\documentclass[12pt]{article}
\begin{document}
\title{On Exact Summations in Long-Range Interactions}
\author{Sergio Curilef \\
\em Departamento de F\'\i sica, Universidad Cat\'olica del Norte, \\Av. Angamos 0610, Antofagasta, Chile \\
scurilef@ucn.cl}
\maketitle
\begin{abstract}
In this work, simple exact results are presented for summations in two-particle potential with
long-range interactions. 
Polygamma function is used to evaluate summations.
Results are found when a periodic media is consider. Periodic boundary conditions 
are applied by symmetric repetitions of a central cell.
Contributions over all space are used in the problem.  The potential depends
strongly on the size of the system; however, forces are convergents anywhere. 
\end{abstract}
\section*{Introduction}

The thermodynamical extensivity imposes short-range interactions
in classical systems. Standard theoretical and experimental behaviors have
been discussed in details since several 
decades~\cite{fisher,Ruelle}.
However, anomalous behavior is obtained when the potential attractive tail behaves very slow.
In recent years, much attention has been paid to physical systems with 
microscopic long-range  interactions ({\em i.e.,} see \cite{SLaAJP99,SCannas} and references therein).

The most frequently applied way to discuss this type of systems is the
Ewald method{\cite{Allen-T}}, 
where neutralizing counter charges are introduced to ensure
convergence of the energy between one particle and the infinite 
replications of other particle. A more recent approach is the Lekner 
method~\cite{JLePhA176,JLePhA157},
where the symmetry of the lattice of identical computational cells is used
to guarantee the convergence of the relevant interaction force.
Here, no neutralizing charges are considered and therefore, we adopt a 
Lekner-like procedure to evaluate the resulting interactions.
We get in all cases (short and long-range 
interactions) convergent thermodynamical quantities. We discuss 
the most appropriate size of the system to evaluate such quantity.

We have a central computational cell in one dimension with normalized 
size $L=1$, this is $-1/2<z<1/2$ where $z$  is the variable of the position. 
This work is an analytical extension of a computational one introduced in~\cite{SCuIJMPC11}.
Systems in one dimension are very important due to effects about ferromagnetism in 
one-dimensional monoatomic metal chains which has been reported recently\cite{PGaNat416}.
Periodic boundary conditions are computed by repetition
of the central cell to infinity. Particles interact with a potential given by
\begin{equation}
v(z) = -J_0 \sum_{k=-\infty}^{\infty} g(|k+z|),
\label{pot1}
\end{equation}
where $J_0$ is the hopping parameter and the summation on $k$
represents all contributions over replicated images. 
Certainly, forces are obtained by $\vec{f}(z) = -\hat{z}dv(z)/dz $.

Now, summations can be written in the following manner
\begin{equation}
\sum_{k=-\infty}^{\infty} g(|k+z|)= \sum_{k=1}^{\infty} g(k-z)
+ g(|z|) +\sum_{k=1}^{\infty} g(k+z).
\label{gs}
\end{equation}
In the present work, the following expressions for potentials are taken into account 
\begin{equation}
g(z) =
\left\{ \begin{array}{cl}
  1/|z| & \mbox{a power law interaction}\\
  \log(|z|)& \mbox{a logarithmic interaction.}
  \end{array}\right.
  \label{g}
\end{equation}
Replacing Eq.(\ref{g}) in  Eq.(\ref{pot1}), $v(z)$ is shown to be divergent; 
then, it is necessary to propose an alternative way for discussing this kind of potential. 
In present calculations, a particular summation (of $1/(k+z)$ on 
$k=1,2,3,...$) is involved in the computation of potentials or forces .
Let us use the definition of the digamma function\cite{GArfken}
\begin{equation}
{\cal F}(z) =  -\gamma -\sum_{k=1}^{\infty} \left( \frac{1}{k+z} -\frac{1}{k} \right)
\label{dig}
\end{equation}
where $\gamma=-{\cal F}(0)= 0.577215664901...$  that has been computed by several methods and 
accuracy\cite{DKnmc16,DSwMC17}. We note that summations are divergents what makes necessary to take a cutoff 
Now, we are going to approximate the summation from Eq.(\ref{gs}) by using the following approximated definition
\begin{equation}
 \sum_{k=1}^{M} \frac{1}{k+z} = {\cal F}(M) -{\cal F}(z)
 \label{FF}
\end{equation}
 where $M\rightarrow\infty$. We verify from Eq.(\ref{dig}) that if $z=M$ is an integer, we can get
\begin{equation}
 {\cal F}(M) = \sum_{k=1}^{M} \frac{1}{k} -\gamma 
 \label{FM}
\end{equation}
which converges only for finite values of $M$ and diverges when $M$ is infinite.
In this way, $M$ provides a natural cutoff for divergent summations.

\section{$1/z$ potential}
From Eq.(\ref{pot1}), the potential can be written as follows
\begin{equation}
v(z) = -J_0\left( \frac{1}{|z|} + \sum_{k=1}^{\infty} \frac{1}{k+z} + \sum_{k=1}^{\infty} \frac{1}{k-z} \right)
\label{vv1}
\end{equation}
and we find a close expression for the two-particle potential with a 
cutoff given by $M$ replications in each direction of the lattice. 
If we replace Eq.(\ref{FF}) in Eq.(\ref{vv1}), we can obtain an exact expression for the potential
\begin{equation}
v(z)   = -J_o \left( \frac{1}{z} + 2{\cal F}(M) -{\cal F}(z)-{\cal F}(-z) \right). 
\label{vz}
\end{equation}
from this equation, we can see that the potential $v(z)$ depends on $M$. 
To eliminate that dependency, we will proceed to scale the potential as it has been
conjectured~\cite{PJuPRB52}. 
For example,
in this particular case, we can define a new scaled variable as follows
\begin{equation}
u(z)   = \frac{v(z)}{{\cal F}(M)+ \gamma } 
\end{equation}
This kind of scaling preserves the Legendre structure and it has been largely 
discussed in recent literature ({\em i.e.,} see \cite{SCuIJMPC11,SCuPLA299,SCuPLA99} and references therein).
In Figure 1(a) we depict the $v(z)/J_0$ as a function of z. 
In this way, from Eq.(\ref{vz}) we verified that the potential does not saturate when $M$ increases. 
This is a typical behavior of a system with long-range interactions. 
Now, in figure 1(b) is depicted the scaled potential $u(z) /J_0$ as a function of z, and we observe 
if $M$ increases, the variable converges. The curve for M=1 (dashed line) is far from the curve for $M=10$ 
(long-dashed line), but as it is expected the curve for $M=10$ is close to $M=100$ (solid line), and we note a
tendency to saturate when $M$ increases.

In addition to this, we can obtain the force between two particles and its replicated images as follows
\begin{equation}
\vec{f}(z)=\hat{z}\frac{d}{dz} v(z)= \hat{z}J_o \left( \frac{1}{z^2} +{\cal F'}(z)-{\cal F'}(-z) \right),
\label{force}
\end{equation}
where 
\begin{equation}
{\cal F}^{(m)}(z) = d^m{\cal F}(z) / dz^m.
\label{poly}
\end{equation}
Polygamma function is obtained by differentiating the digamma function repeatedly. 
In the force calculation from Eq.(\ref{force}), no dependency on $M$ is obtained; 
this was appointed in a previous computational test\cite{SCuIJMPC11}. 
In general, we use the symmetry of the lattice 
and contributions on a particle from images of the other one.  

The simplest case is one particle in the central cell and the potential must consider
the potential on the central particle from its replicated images. 
This is therefore explicitly given by the following sum, 
\begin{equation}
v   = -2 J_o  \sum_{k=1}^{M} \frac{1}{k} ,
\label{vv}
\end{equation}
and using Eq.(\ref{FM}) in the potential $v$, we can get
\begin{equation}
v   = -2 J_o  \left( {\cal F}(M) + \gamma \right).
\end{equation}
Also, it is possible to see that the force is zero, this is $\vec{f} = \hat{k} dv/dk = 0$,
for the simple case of one central particle in a linear system that only interacts with its images. 

\section{Logarithmic potential}
When the potential is logarithmic, the force is wrote as follows
\begin{equation}
\vec{f}(z)= \hat{z}J_0\sum_{k=-\infty}^{\infty} \frac{1}{k+z}  
= \hat{z}J_0\left( \frac{1}{z} + \sum_{k=1}^{\infty} \frac{1}{k+z}  - 
\sum_{k=1}^{\infty} \frac{1}{k-z} \right)
\end{equation}
By using the identity 
\begin{equation}
\cot(z)= \frac{1}{\pi}\sum_{k=-\infty}^{\infty} \frac{1}{k+z/\pi},
\end{equation}
we can obtain an additional relation for representing an exact form for the force when the present
logarithmic potential is considered, this is 
\begin{equation}
\vec{f}(z)= \hat{z}J_o \pi \cot(\pi z)
= \hat{z}J_o \left( \frac{1}{z} -{\cal F}(z)+{\cal F}(-z) \right).
\end{equation}
The advantage for the second one is the explicit contribution to the force. The first
term is the interaction on one particle due to the other particle in a particular cell, 
the second (and third) term is the contribution due to the images of the second particle situated on 
the right (left) of the central cell.

\section{Discussion}
On one hand, we have obtained a close expression for summations with the long-range interactions. 
Polygamma function
is used and evaluation is fast and rapidly convergent.
Potentials are strongly dependent on the size of systems.  
From Eq.(\ref{force}), it is shown that the forces do not depend on the size of the system, 
such as it was obtained in a previous numerical work\cite{SCuIJMPC11}. 
A direct cutoff is provided for potential summations by number $M$ of repetitions of the 
central cell over all space. 

On the other hand, in the short range interaction regime when the attractive 
tail falls very fast as  $1/z^{1+m}$ law, the potential is given by
\begin{equation}
v(z) = \sum_{k=-\infty}^{\infty} \frac{1}{|k+z|^{1+m}}
\end{equation}
According to Eq.(\ref{poly}) we can obtain an explicit definition for polygamma function, given by
\begin{equation}
{\cal F}^{(m)}(z) = (-1)^{m+1} m! \sum_{k=1}^{\infty}\frac{1}{(k+z)^{1+m}},
\end{equation}
where $z$ is a real variable. This relation allows us a general expression for potential with the present
power law short-range interactions.
\begin{equation}
v(z) = -J_0\frac{(-1)^{m+1}}{m!}\left(\frac{(-1)^{m+1}}{z^{1+m}} +{\cal F}^{(m)}(z) +{\cal F}^{(m)}(-z) \right),
\end{equation}
being $m=1,2,3,...$. In this result is shown that $v(z)$ does not depend on size of the system 
when interactions are short ranged. Different terms are in relation to several contribution due to 
inter-particle interaction and contributions due to replicated images in each directions.

Finally, this kind of generalizations are added to several contribution in this area, and the problem related
to periodic boundary conditions is a topic for several studies. 
Thermodynamical limit for infinite systems with finite range of interactions is very well defined. 
However, as it has been shown, 
if the range of interactions is greater than the size of the system, thermodinamical quantities
depend strongly on the size of system. It is very important to understand the microscopic behavior
of the particles because the average from computer simulations comes from the average of finite number
of particles in systems. Advances on this topic will give us a view to understand the thermodynamical 
behavior for this kind of systems.

\section*{Acknowledgments}
 This work has received partial support by FONDECYT, grant 1010776.

\newpage
\centerline{ FIGURE CAPTION}

Figure 1 (a) Potential does not converge when M increases. (b) Scaled potential shows a tendency to saturate 
when $M$ increases. Curves in the figure are depicted for $M=1$ (dashed line) , $M=10$ (long-dashed line) 
and $M=100$ (solid line).


\begin{thebibliography}{88}
\bibitem{fisher} M. E. Fisher, Arch. Rat. Mech. An. {\bf 17} (1964) 377.
\bibitem{Ruelle} D. Ruelle, {\em "Statistical Mechanics"}, Imperial College Press and World Scientific (1999).
\bibitem{SLaAJP99} S. K. Lamoreaux, Am. J. Phys. {\bf 67}, 850-861 (1999). 
\bibitem{SCannas} S. A. Cannas, C. M. Lapilli, C. A. Stariolo, cond-mat 0307163.
\bibitem{Allen-T} M. P. Allen and D.J. Tildesley, {\it Computer Simulations 
in Liquids}, (Oxford
Science Publications, 1994).
\bibitem{JLePhA176} J. Lekner, Physica {\bf A 176}, 485 (1991)
\bibitem{JLePhA157}  J. Lekner, Physica {\bf A 157}, 826 (1989)
\bibitem{SCuIJMPC11} S. Curilef, Int. J. Mod. Phys. {\bf C 11} (2000) 629. 
\bibitem{PGaNat416} P. Gambardella, A. Dallmeyer, K. Maiti, M.C. Malagoli, W. 
Eberhardt, K. Kern, and C. Carbone, Nature {\bf 416} (2002) 301
\bibitem{GArfken} G. B. Arfken and H. J. Weber, {\em "Mathematical methods for physicists"}, 
Fourth Edition Academic Press (1995).
\bibitem{DKnmc16} D. E. Knuth, Math. Comput. {\bf 16} (1962) 275;
\bibitem{DSwMC17} D. W. Sweeney  Math. Comput. {\bf 17} (1963) 160.
\bibitem{PJuPRB52} P. Jund, S. G. Kim and C. Tsallis, Phys. Rev. {\bf B 52} (1995) 50.
\bibitem{SCuPLA299} S. Curilef, Phys. Lett. {\bf A 299} (2002) 366.
\bibitem{SCuPLA99} S. Curilef and C. Tsallis, Phys. Lett. {\bf A 264} 270 (1999);  
\end{thebibliography}
\end{document}